\documentclass[11pt,a4paper]{article}
\RequirePackage[english]{babel}
\RequirePackage[T1]{fontenc}
\RequirePackage{graphicx}
\usepackage{amsmath}
\usepackage{hepnames}

\begin{document}

\title{\begin{flushright}
Nikhef 2013-037  
\end{flushright} 
\vspace{10mm}
An extended Heitler-Matthews model for the full hadronic cascade in cosmic air showers.}
\author{J.M.C. Montanus \\
\\ \\Nikhef}
\date{March 5, 2014}
\maketitle

\begin{abstract}
The Heitler-Matthews model for hadronic air showers will be extended to all the generations of electromagnetic subshowers in the hadronic cascade. The analysis is outlined in detail for showers initiated by primary protons. For showers initiated by iron primaries the part of the analysis is given for as far as it differs from the analysis for a primary proton. Predictions for shower sizes and the depth of maximum shower size are compared with results of Monte Carlo simulations. The depth of maximum as it follows from the extrapolation of the Heitler-Matthews model restricted to the first generation of electromagnetic subshowers is too small with respect to Monte Carlo predictions. It is shown that the inclusion of all the generations of electromagnetic subshowers leads to smaller predictions for the depth of maximum and to smaller predictions for the elongation rate. The discrepancy between discrete model predictions and Monte Carlo predictions for the depth of maximum can therefore not be explained from the number of generations that is taken into consideration. An alternative explanation will be proposed.
\end{abstract} 

\newpage
\section{Introduction}
A simplified description of the longitudinal evolution of electromagnetic showers is given by the Heitler model \cite{Heitler}. Starting with a primary particle of energy $E_0$, the number of particles doubles every splitting length $d=\lambda_r \ln2$, where the radiation length $\lambda_r$ is about 37 g cm$^{-2}$. The doubling stops when the energy per particle is equal to the critical energy $\xi_c^e \approx 85$ MeV. The resulting Heitler profile is
\begin{eqnarray}\label{0}
N(X)=\left\{ \begin{array}{ll} 2^{X/d}  & \ , \ X 
\leq n_c^{e} d \ ; \\ 0 & \ , \ X > n_c^{e} d \ , \end{array} \right.
\end{eqnarray}
where $n_c^{e}$ is maximum the number of steps: $n_c^{e} \ln 2=\ln (E_0/ \xi_c^e )$.
\\ \\
A Heitler model for the hadronic cascade in air showers has been constructed by Matthews \cite{Matthews2}. The Heitler-Matthews model is useful for the explanation of hadronic cascades as well as for the analytical derivation of relations between quantities as primary energy, muon number, electron number and depth of maximum shower size \cite{Horandel, Ulrich2, Engel}. For the prediction of the number of charged particles it is assumed that each hadronic interaction results in $M_\textrm{ch}=10$ charged pions and $\frac{1}{2}M_\textrm{ch}=5$ neutral pions. That is, the total multiplicity $M$ is equal to 15. The neutral pions initiate electromagnetic subshowers when they decay into photons. For the prediction for the depth of maximum shower size, restricted to the first generation of electromagnetic subshowers, the multiplicity and interaction length are parameterized by the energy of the interaction. 
\\ \\
The atmosphere is divided into layers of atmospheric thickness $d_\textrm{I}$. After the traversing of each layer the number of charged pions is assumed to be $M_\textrm{ch}$ times larger if $d_\textrm{I}=\lambda_\textrm{I} \ln 2$, where $\lambda_\textrm{I} = 120$ g cm$^{-2}$ is the interaction length of strongly interacting pions. Consequently, after $n$ layers the number of charged pions is $\left( M_\textrm{ch}\right)^n$. The energy per pion is
\begin{equation}\label{1}
E_{\pi,n} = \frac{E_0 }{ M^n} \ .
\end{equation}
The stopping energy is estimated on the basis of the finite lifetime of the pions in the atmosphere. For this it suffices to consider the approximate relation between atmospheric depth and height: 
\begin{equation}\label{2}
X(h)=1030 \cdot e^{-h/8} \ \leftrightarrow \ \ \ h(X)=8\ln (1030/X) \ ,
\end{equation}
where $X$ is the depth in g cm$^{-2}$ and $h$ is the height in km. Neutral pions decay almost immediately into two photons, $c \tau =25$ nm \cite{PDG}. Each resulting photon starts an electromagnetic shower. The decay length of the charged pions is $\gamma c \tau$, where $c \tau =7.8$ m \cite{PDG}. The decay length is of the order of a kilometer because of the relativistic time dilation. As a consequence charged pions may interact with the atmosphere and propagate the hadronic shower, before decay. If the probability for decay in the next layer is larger than the probability of a hadronic interaction, the pions are assumed to decay and the cascade stops. This happens after $n_c$ layers. The corresponding energy of the decaying charged pions, the stopping energy $\xi_ c^\pi$, follows from
\begin{equation}\label{3}
\xi_c ^\pi =\frac{E_0 }{ M^{n_c}} \ .
\end{equation}
The stopping energy turns out to be around 20 GeV.

\section{Model parameters}

\begin{figure}[htbp]
\begin{center}
\includegraphics[width=9.0cm]{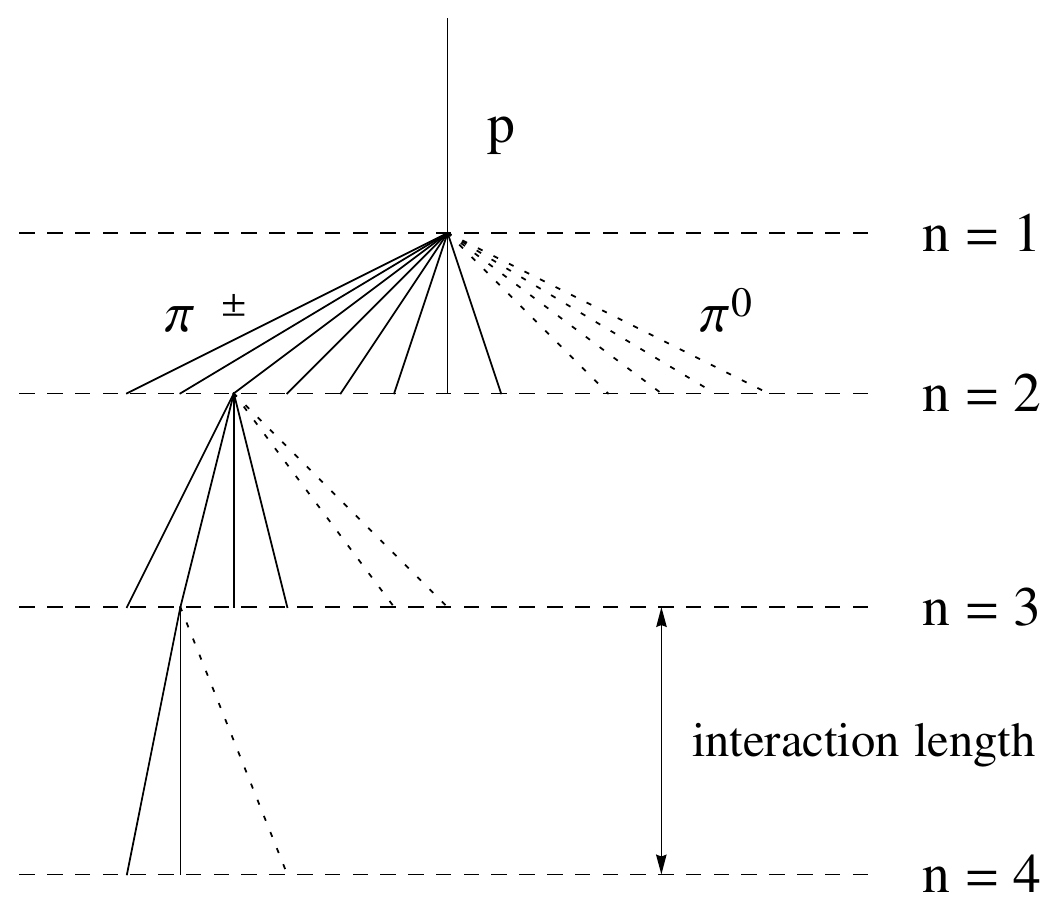}
\caption{The hadronic cascade for energy dependent interaction lengths.}
\label{scheme}\end{center}
\end{figure}

In this paper the Heitler-Matthews model is extended to all the generations of pions in the shower. The complete analysis will be improved by consequently taking the multiplicity $M$ and interaction length $\lambda_\textrm{I}$ to depend on the energy of the hadron in the shower. One of the consequences is that the thickness of the cascade layers increases with depth, see Fig. \ref{scheme}.
\\
We will take the energy dependence of the $\Ppi$-air multiplicity and interaction length to be given by Monte Carlo event generators based on QCD and parton models. The calculated $\Ppi$-air charged multiplicity, see Fig. 5 of \cite{Roesler}, Fig. 7 of \cite{Alvarez} and Fig. 5 of \cite{Knapp}, suggests the relation
\begin{equation}\label{4}
M_\textrm{ch} \approx 0.1 \cdot {E}^{0.18} \ ,
\end{equation}
where $E$ is the energy in eV. Taking the ratio of the charged and neutral pions as $2:1$, we have for the total multiplicity
\begin{equation}\label{5}
M \approx 0.15 \cdot {E}^{0.18} \ ,
\end{equation}
It should be emphasized that the relation between multiplicity and energy is rather uncertain since different models predict different multiplicities. In particular for large energies the differences can be large, even more than 100\%.  From Fig. 5 of \cite {Roesler} and Fig. 7 of \cite{Alvarez} we see that the pion multiplicity does not differ substantially from the proton multiplicity. We therefore will use the relations (\ref{4}) and (\ref{5}) for both the multiplicity in proton-air ($\Pproton$-air) and pion-air ($\pi$-air) interactions. A parameterization with other values for the constants will, of course, affect the results quantitatively. It does, however, not alter the results qualitatively.
\\ \\
Both the $\Pproton$-air and $\pi$-air production cross sections grow with energy.  The \Pproton-air production cross section at large energies obtained from observations of extensive air showers are in good agreement with QGSJET predictions \cite{Ulrich, Abreu, Aglietta,Aielli}. For the present analysis we will therefore use the QGSJET predictions for the $\Pproton$-air production cross section. We will also use the QGSJET predictions for the $\pi$-air production cross section \cite{Alvarez, Stanev}. From these cross sections approximations for the energy dependent interaction lengths can be derived which are sufficiently accurate for our purpose. For $\pi$-air this is:
\begin{equation}\label{6}
\lambda_{\pi \textrm{-air}} \mbox{  [g cm}^{-2}] \approx 200-3.3 \ln (E [\textrm{eV}])\ . 
\end{equation}
For $\Pproton$-air this is  
\begin{equation}\label{7}
\lambda_\textrm{p-air} \mbox{  [g cm}^{-2}] \approx 145-2.3 \ln (E [\textrm{eV}]) \ .
\end{equation}
For the hadronic cascade Matthews and H\"orandel take $d_\textrm{I}=\lambda_\textrm{I} \ln 2$ as the relation between layer thickness and interaction length\cite{Matthews2,Horandel}. This might have been motivated by the expression $\lambda_r \ln 2$ for the splitting length in the electromagnetic cascade. There the ratio $\ln 2$ results from the translation of the radiation length to the splitting length. In an intermediate model for electromagnetic showers the splitting length, $\lambda_r \ln 2$, is effectively used as the electromagnetic interaction length \cite{Montanus}. For the hadronic cascade, however, there is no reason for the ratio $\ln 2$ since $\lambda_\textrm{I}$ is already an interaction length. As a consequence the thickness of the interaction layer in hadronic cascades is equal to the interaction length. For hadronic showers we will therefore use the relation $d_\textrm{I}=\lambda_\textrm{I}$.   

\section{The hadronic cascade for a primary proton}
Now we consider a hadronic cascade where the hadronic particles interact after having traversed a layer of atmosphere. The thickness of each layer will be taken equal to the actual interaction length as given by (\ref{6}) or (\ref{7}). After each interaction $M$ pions are produced as given by (\ref{5}). In accordance with the Heitler model for electromagnetic showers, the energy is assumed to be equally divided over the particles produced. After each interaction the new energy of the charged hadrons then follows from a successive application of the equation
\begin{equation}\label{20}
E_{j+1}=\frac{E_j}{M(E_j)}  \ .
\end{equation}
Starting with a primary proton with energy $E_0$ the energy of the particles after the first interaction is  
\begin{equation}\label{21}
E_1= \frac{E_0}{0.15 \cdot E_0^{0.18}} \approx 6.7 \cdot E_0^{0.82} \ .
\end{equation}
After the second interaction this is 
\begin{equation}\label{22}
E_2= \frac{E_1}{0.15 \cdot E_1^{0.18}} \approx 6.7 \cdot E_1^{0.82} \approx  6.7^{1.82} \cdot E_0^{0.82^2} \ .
\end{equation}
Repeating the iteration we find for the energy per particle after $n$ interactions
\begin{equation}\label{23}
E_n=  6.7^{\alpha_n} \cdot E_0^{\beta_n} \ ,
\end{equation}
where
\begin{equation}\label{24}
\alpha_n = \frac{1-0.82^n}{1-0.82} \ , \ \ \ \beta_n = 0.82^n \ .
\end{equation}
For the interaction lengths we obtain for the primary proton, $n=0$,
\begin{equation}\label{25}
\lambda_0 = \lambda_\textrm{p-air}(E_0) = 145- 2.3 \ln (E_0)
\end{equation}
and for the produced pions, $n \geq 1$,
\begin{equation}\label{26}
\lambda_n = \lambda_{\pi \textrm{-air}}(E_n)= 200- 3.3 \ln (E_n) \ .
\end{equation}
With the substitution of (\ref{23}) this is 
\begin{equation}\label{27}
\lambda_n = 200-6.3 \alpha_n -3.3 \beta_n \ln (E_0) \ , \ \ \ n \geq 1 \ .
\end{equation}
The atmospheric depth of the $n_c$-th interaction is given by
\begin{equation}\label{28}
X(n_c) =  \sum_{n=0}^{n_c-1} \lambda_n \ ,  \ \ \ n_c \geq 1   \ .
\end{equation}
Substitution of (\ref{25}) and (\ref{27}) leads to
\begin{equation}\label{29}
X(n_c) =  200 n_c -55 -6.3 \sum_{n=1}^{n_c-1} \alpha_n - \left( 2.3+3.3 \sum_{n=1}^{n_c-1}\beta_{n} \right)\ln (E_0) \ .
\end{equation}
With the approximate relation (\ref{2}) between height and atmospheric depth, we obtain for the difference in height between the $n_c$-th and $(n_c+1)$-th interaction:
\begin{equation}\label{30}
\Delta h [\textrm{km}] =  8 \ln \left( \frac{X(n_c+1)}{X(n_c)} \right)  \ .
\end{equation}
For the decay length of the charged pions after the $n_c$-th interaction we find
\begin{equation}\label{31}
c \tau \gamma [\textrm{km}] \approx 7.8 \cdot 10^{-3} \frac{E_{n_c}}{m_{\pi \pm}} \approx 5.6 \cdot 10^{-11} \cdot 6.7^{\alpha_{n_c}} \cdot E_0^{\beta_{n_c}} \ , 
\end{equation}
where we have taken $140$ MeV/c$^2$ for the mass of the charged pions \cite{PDG}. We will follow Matthews with the reasonable assumption that the pions will decay when the decay length is half the layer thickness: $\gamma c \tau \approx \frac{1}{2} \Delta h$ \cite{Matthews2}. That is 
\begin{equation}\label{32}
5.6 \cdot 10^{-11} \cdot 6.7^{\alpha_{n_c}} \cdot E_0^{\beta_{n_c}} =  4 \ln \left( \frac{X(n_c+1)}{X(n_c)} \right) \ , 
\end{equation}
In this equation we substitute integer values for $n_c$ and solve numerically for the primary energy $E_0$. Results of interest are shown in table 1, 2 and 3. In table 1 the height $h$ is calculated with Eq. (\ref{2}). In table 2 and 3 the interaction lengths respectively multiplicities are given for all interactions up to the final one after which decay occurs.  Of course, the number of interactions in the cascade and thus also the penetration depth of the shower increases with the energy of the primary particle. Since a larger atmospheric depth at lower altitudes corresponds to a larger  probability for a charged pion to interact before decay, we expect a smaller stopping energy. From the last entry in table 1 we see the stopping energy indeed decreases for increasing primary energy. 
\begin{table}[htbp]
\renewcommand{\arraystretch}{1.5}
\setlength{\tabcolsep}{6pt}
\begin{tabular}{|c |c |c |c |c |c |c |c |c |}\hline 
$n_c$ & $E_0$ [eV] & $X(n_c)$ & $X(n_c+1)$ & $h(n_c)$ [km] & $\Delta h$ [km] & $\xi_c^{\pi}$ [eV]   \\ \hline 
$1$ & $1.5 \cdot 10^{12}$ & $81$ & $198$ & $20.4$ & $7.2 $ & $6.4 \cdot 10^{10}$  \\ 
$2$ & $2.9 \cdot 10^{13}$ & $184$ & $303$ & $13.8$ & $4.0 $ & $3.6 \cdot 10^{10}$  \\ 
$3$ & $1.4 \cdot 10^{15}$ & $275$ & $396$ & $10.6$ & $2.9 $ & $2.6 \cdot 10^{10}$  \\ 
$4$ & $1.9 \cdot 10^{17}$ & $352$ & $473$ & $8.6$ & $2.4 $ & $2.1 \cdot 10^{10}$  \\ 
$5$ & $8.2 \cdot 10^{19}$ & $409$ & $531$ & $7.4$ & $2.1 $ & $1.9 \cdot 10^{10}$ \\ \hline
\end{tabular}
\caption{Characteristics of hadronic cascades for a primary proton.}
\end{table}
\begin{table}[htbp]
\renewcommand{\arraystretch}{1.5}
\setlength{\tabcolsep}{5pt}
\begin{tabular}{|c |c |c |c |c |c |c |c |c |}\hline 
$n_c$ & $E_0$ [eV] & $\lambda_0$ & $\lambda_1$ & $\lambda_2$ & $\lambda_3$ & $\lambda_4$ & $\bar \lambda$ \\ \hline
$1$ & $1.5 \cdot 10^{12}$ & $81$ & -- & -- & -- & -- & 81 \\ 
$2$ & $2.9 \cdot 10^{13}$ & $74$ & $110$ & -- & -- & -- & 92 \\ 
$3$ & $1.4 \cdot 10^{15}$ & $65$ & $99$ & $111$ & -- & -- & 92 \\ 
$4$ & $1.9 \cdot 10^{17}$ & $53$ & $86$ & $100$ & $112$ & -- & 88 \\ 
$5$ & $8.2 \cdot 10^{19}$ & $40$ & $70$ & $87$ & $101$ & $112$ & 82 \\ \hline
\end{tabular}
\caption{Subsequent interaction lengths in hadronic cascades for a primary proton. The $\lambda_n$ and the cascade average $\bar \lambda$ are in g cm$^{-2}$.}
\end{table}
\begin{table}[htbp]
\renewcommand{\arraystretch}{1.5}
\setlength{\tabcolsep}{5pt}
\begin{tabular}{|c |c |c |c |c |c |c |c |c |}\hline 
$n_c$ & $E_0$ [eV] & $M_0$ & $M_1$ & $M_2$ & $M_3$ & $M_4$ & $\bar M$ \\ \hline
$1$ & $1.5 \cdot 10^{12}$ & $23$ & -- & -- & -- & -- & $23$ \\ 
$2$ & $2.9 \cdot 10^{13}$ & $40$ & $20$ & -- & -- & -- & $28$\\ 
$3$ & $1.4 \cdot 10^{15}$ & $80$ & $36$ & $19$ & -- & -- & $38$ \\ 
$4$ & $1.9 \cdot 10^{17}$ & $193$ & $75$ & $34$ & $18$ & -- & $55$ \\ 
$5$ & $8.2 \cdot 10^{19}$ & $576$ & $183$ & $72$ & $33$ & $18$ & $85$ \\ \hline
\end{tabular}
\caption{Subsequent multiplicities and their geometric mean in hadronic cascades for a primary proton.}
\end{table}
\\ \\
The number of muons is given by
\begin{equation}\label{32a}
N_{\mu} = \left( \frac{2}{3} \right)^{n_c} \cdot \prod_{n=0}^{n_c-1}  M_n \ . 
\end{equation}
For a primary proton with energy $1.4 \cdot 10^{15}$ eV, as an example, the number of muons is about $1.6 \cdot 10^4$. The energy of the final pions is about $2.6 \cdot 10^{10}$ eV, see last entry of table 1. For the quantity 
\begin{equation}\label{33}
\beta = \frac{ \ln N_\mu}{\ln \left( E_0 / \xi_c^\pi \right)} 
\end{equation}
we then obtain the value $0.89$. For other primary energies the value of $\beta$ is found to increase slightly from 0.88 through 0.92 for a primary energy increasing from $10^{12}$ eV through $10^{20}$ eV. Since we work with larger multiplicities, these values are slightly larger than the one obtained by Matthews \cite{Matthews2}. 
\\ \\
From the last column in table 3 we find the following approximate relation between the effective multiplicity (the geometric mean multiplicity) and the primary energy: $\ln \bar M \approx 1.05 +  0.074 \ln E_0$. The effective charged multiplicity then is given by $\ln \bar M_{\textrm{ch}} \approx 0.65 +  0.074 \ln E_0$. Substituting the latter in the Matthews' expression \cite{Matthews2}  
\begin{equation}\label{34}
\beta = 1 - \frac{\kappa}{3 \ln M_{\textrm{ch}}} \ ,
\end{equation}
where $\kappa$ is the inelasticity, we obtain a refinement for the energy dependence:
\begin{equation}\label{34b}
\beta = 1 - \frac{\kappa}{1.9+0.22 \ln E_0} \ .
\end{equation}

\section{Hybrid Heitler scheme}
\begin{figure}[htbp]
\begin{center}
\includegraphics[width=11.0cm]{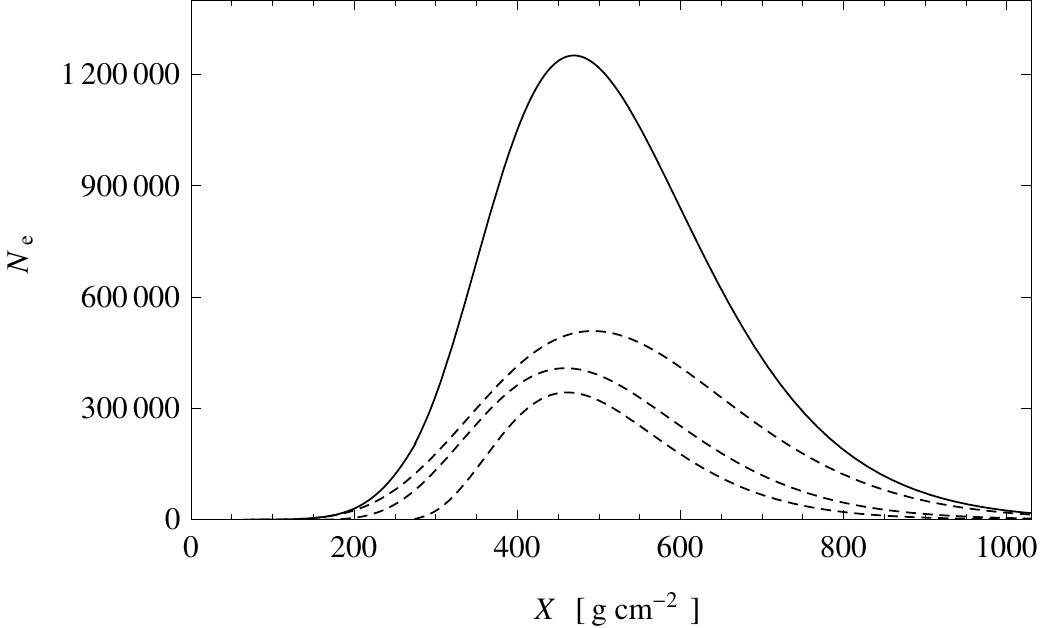}
\caption{Total electromagnetic shower profile (solid) as it results from the addition of the profiles of the electromagnetic subshowers (dashed) for a 1.4 PeV proton primary.}
\label{profile1}
\end{center}
\end{figure}

\begin{figure}[htbp]
\begin{center}
\includegraphics[width=11.0cm]{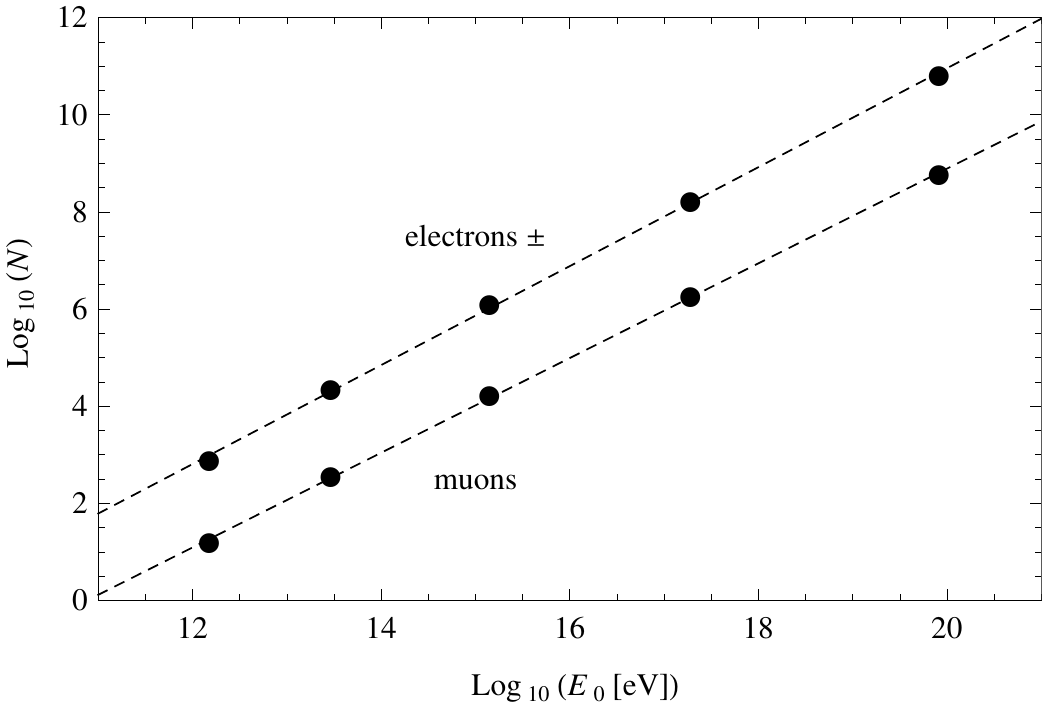}
\caption{The maximum number of electrons ($\pm$) and muons as a function of the energy of the primary proton (dots) and the linear fits (dashed).}
\label{sizes}
\end{center}
\end{figure}

\begin{figure}[htbp]
\begin{center}
\includegraphics[width=11.0cm]{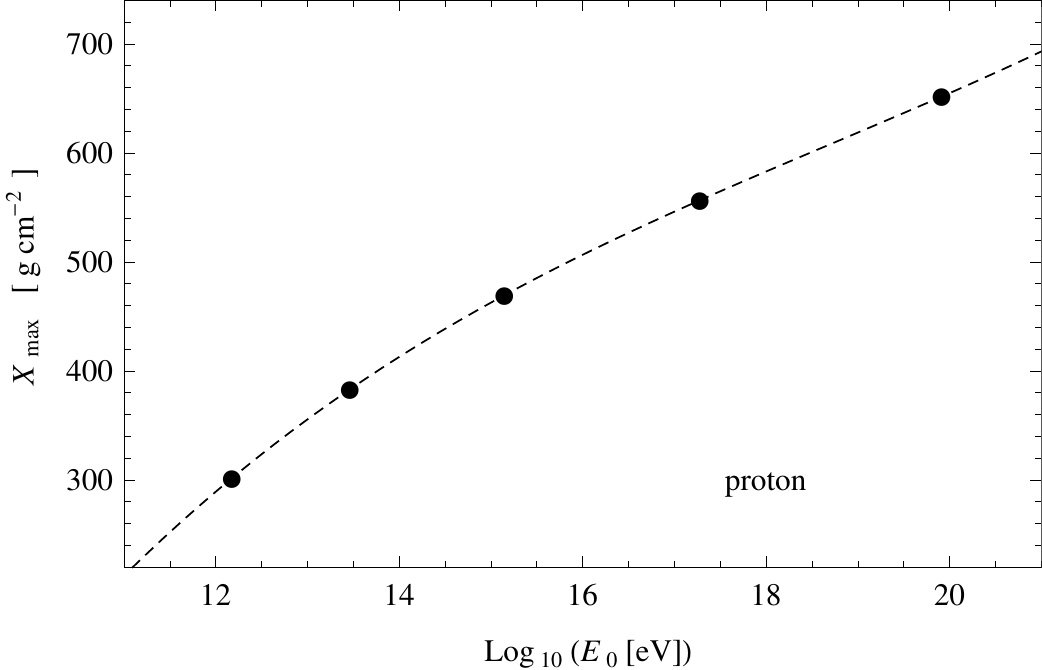}
\caption{The depth of maximum shower size as a function of energy of a proton primary for the situation with complete inelasticity (dots) and the corresponding cubic fit (dashed).}
\label{proxmax}
\end{center}
\end{figure}

The Heitler line of reasoning can also be applied to electromagnetic shower profiles other than the Heitler profile (\ref{0}). Let $N(X;E)$ be an electromagnetic shower profile which for a primary energy $E$ has its maximum $N_\textrm{max} (E)$ at depth $X_\textrm{max}(E)$. For twice the primary energy the particle will split into two particles with energy $E$ after one splitting length $d=\lambda_r \ln2$. The corresponding shower profile can be regarded as twice the shower profile for $E$ shifted with $d$ towards a larger depth. As a consequence $N_\textrm{max} (2E)=2N_\textrm{max} (E)$ and $X_\textrm{max}(2E)=X_\textrm{max}(E)+d$. The latter corresponds to the  following elongation rate: 
\begin{equation}\label{35}
\frac{\textrm{d} X_\textrm{max}}{\textrm{d} \log_{10} E} \approx \frac{\Delta X_\textrm{max}}{\Delta \log_{10} E} = \frac{d}{ \log_{10}2} = \lambda_r \ln 10 \ .
\end{equation}
This scheme will be applied to the present hadronic cascades. Each time when neutral pions decay into photons an electromagnetic subshower is initiated and the corresponding longitudinal profile is substituted. It is similar to what is done in the hybrid Monte Carlo model CONEX \cite{CONEX}. For the electromagnetic shower profile we take the Greisen function \cite{Gaisser}:
\begin{equation}\label{36}
N_e (X)=\frac{0.31}{\sqrt{y_c}} \cdot e^{\frac{X}{\lambda_r}(1-1.5 \ln s )} \ ,
\end{equation}
where $y_c=\ln (E_0/\xi_c^e)$ and where
\begin{equation}\label{37}
s=\frac{3X}{X+2X_\textrm{max}} 
\end{equation}
is the age parameter. For showers initiated by a photon a good prediction for the depth of maximum is given by
\begin{equation}\label{38}
X_{\textrm{max},\gamma} \mbox{  [g cm}^{-2}] = n_c \cdot d = y_c \cdot \lambda_r \approx 85  \log_{10} E_0 - 675 \textrm{ }  \ .
\end{equation}
For energies larger than 1 EeV, the depth of maximum for photon showers is larger than predicted by (\ref{38}) because of the Landau-Pomeranchuk-Migdal (LPM) effect \cite{Landau,Landau2,Migdal}. See for instance the corresponding curve in Fig. 13 of \cite{Rebel}. By means of a hybrid Monte Carlo model the consequences of the LPM effect for hadronic air showers are found negligible for primary energies below $3 \cdot 10^{20}$ eV \cite{Alvarez}. For such extremely high primary energies the large multiplicity causes the energy of the decay photons after the first interaction to be below 1 EeV, outside the LPM regime. We will therefore conveniently take (\ref{38}) for the depth of maximum shower size of electromagnetic (sub)showers. The total electromagnetic shower profile is obtained by adding the profiles of the electromagnetic subshowers. As an example we consider a shower caused by a $1.4 \cdot 10^{15}$ eV primary proton. The first interaction at depth 65 g/cm$^2$ (Table 2) produces an expected number of 80 pions (Table 3). Equal energy division over the secondaries leads to $1.75 \cdot 10^{13}$ eV for each pion. One-third of the pions are assumed to be neutral. Also for the gamma production we will assume equal energy division. Then each neutral pion decays into two gammas with energy $8.8 \cdot 10^{12}$ eV. Gamma showers with this initial energy have a depth of maximum at about 425 g/cm$^2$. The depth of maximum of this electromagnetic subshower is therefore at $490$ g/cm$^2$. The charged pions survive the next interaction layer of 99 g/cm$^2$ to cause a second generation of gammas. In a similar way as for the first generation the depth of maximum of the second and third generation are found to be both about 460 g/cm$^2$. The total energy contents in the three subsequent electromagnetic subshowers are $\frac{1}{3}E_0$, $\frac{2}{9}E_0$ and $\frac{4}{27}E_0$ respectively. After the third interaction the charged pions are assumed to decay into a muon. The contribution of all three generations of electromagnetic showers to the total shower is illustrated in Fig. \ref{profile1}. 
\\ \\
Since the depth of maximum for the second and third generation is smaller than for the first generation, the depth of maximum for the total electromagnetic shower will also be smaller than predicted by the first generation. For the example above: the depth of maximum of the total shower is 470 g/cm$^2$, which is 20 g/cm$^2$ smaller than the value 490 g/cm$^2$ for the first generation. These differences are even larger for larger primary energies.
\\ \\
The depth of maximum and the maximum number of electrons and positrons of the total electromagnetic shower is found by numerical inspection. This semi-analytical approach is utilized to be able to consider all the generations of subshowers. The number of muons is found by means of Eq. (\ref{32a}). In Fig. \ref{sizes} we have plotted the maximum size of the total electromagnetic shower and the number of muons as a function of initial energy.
\\
A linear fit, see the dashed lines in Fig. \ref{sizes}, for the maximum number of electrons ($\pm$) and number of muons yields
\begin{equation}\label{39}
N_\textrm{e}  \approx 0.57 \cdot ( E_0 [\textrm{GeV}] )^{1.019} 
\end{equation}
and
\begin{equation}\label{40}
N_\mu \approx 0.015 \cdot ( E_0 [\textrm{GeV}] )^{0.975} 
\end{equation}
respectively. The values of $N_e + 25 N_\mu$ for different values of the primary proton energy $E_0$ are close to the ones obtained with Monte Carlo models, e.g. Fig. 3 of \cite{Glasmacher}. 
\\ \\
The depth of the maximum shower size of the total electromagnetic shower is also determined by numerical inspection. In Fig. \ref{proxmax} the depth of maximum shower size is plotted as a function of the energy of the primary proton.
\\
A cubic fit, see the dashed line in Fig. \ref{proxmax}, yields
\begin{equation}\label{41}
X_{\textrm{max,p}} \approx -1796 + 314.3 \lg E_0 - 14.9 (\lg E_0)^2 + 0.265 (\lg E_0)^3 \ ,
\end{equation}
where the common notation $\lg$ for $\log_{10}$ is used for abbreviation. The latter expression is the result for a primary proton as indicated by the subscript `p'. It parameterizes the energy dependence of the elongation rate. For energies above $10^{14}$ eV the curve is almost linear:
\begin{equation}\label{42}
X_{\textrm{max,p}} \approx -140 + 40 \lg E_0 \ .
\end{equation}
In this energy region the elongation rate is constant. Clearly, the depth of maximum on the basis of all the generations of gamma showers is small in comparison with Monte Carlo simulations: 40 instead of 58 g cm$^{-2}$ per decade of energy. An analytical estimate can be obtained by considering solely the first generation of $\gamma$'s \cite{Matthews2, Horandel}. Then, with the present model parameters, 
\begin{equation}\label{43}
X_{\textrm{max,p}} \approx \lambda_{\textrm{p-air}} + X_{\textrm{max},\gamma}(E_0/(2M))\approx -485+64.4 \lg E_0 \ .
\end{equation}
That the corresponding elongation rate is close to value predicted by Monte Carlo simulations should be considered as coincidental. At this point it is instructive to consider also the second generation in the analytical estimation. To this end we leave the constants in the expression for the multiplicity unspecified: $M = q \cdot {E}^{p}$. The depth of maximum of the first and second generation of gamma showers then respectively read
\begin{equation}\label{44}
X_{\textrm{max,p}} = \lambda_{\textrm{p-air}}(E_0) + X_{\textrm{max},\gamma}(E_0/(2qE_0^p)) 
\end{equation}
and
\begin{equation}\label{45}
X_{\textrm{max,p}} = \lambda_{\textrm{p-air}}(E_0) + \lambda_{\pi \textrm{-air}}(E_1)+ X_{\textrm{max},\gamma}(E_1/(2qE_1^p)) \ .
\end{equation}
With the substitution of $E_1=E_0/(qE_0^p)$ these expression reduce to
\begin{equation}\label{46}
X_{\textrm{max,p}}  \approx (79.7-85p) \lg E_0 -556-85\lg q 
\end{equation}
respectively
\begin{equation}\label{47}
X_{\textrm{max,p}} \approx (72-162p+85p^2)\lg E_0 -356  +(85p-162)\lg q\ .
\end{equation}
For reasonable values for $p$ and $q$, in the neighbourhood of $p\approx 0.18$ and $q\approx 0.15$, the depth of maximum of the second generation is smaller than the one from the first generation. The inclusion of the second generation in the analysis will therefore decrease the depth of maximum.

\section{Iron primary}

\begin{figure}[htbp]
\begin{center}
\includegraphics[width=11.0cm]{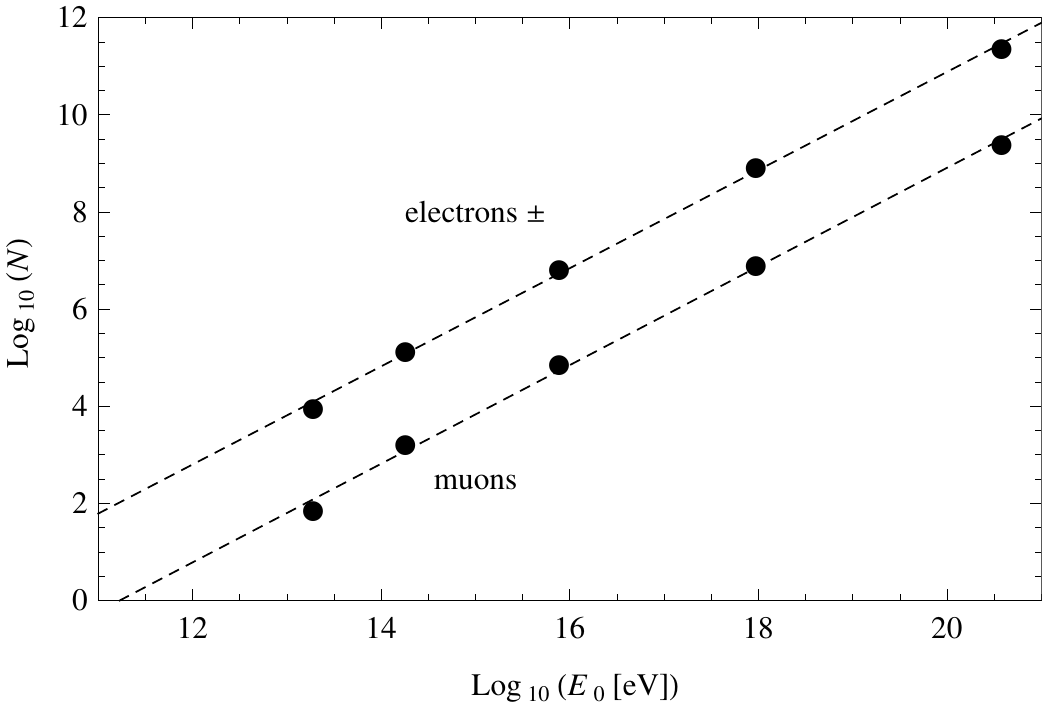}
\caption{The maximum number of electrons ($\pm$) and muons as a function of the energy of the primary iron (dots) and the linear fits (dashed).}
\label{irosizes}
\end{center}
\end{figure}

\begin{figure}[htbp]
\begin{center}
\includegraphics[width=11.0cm]{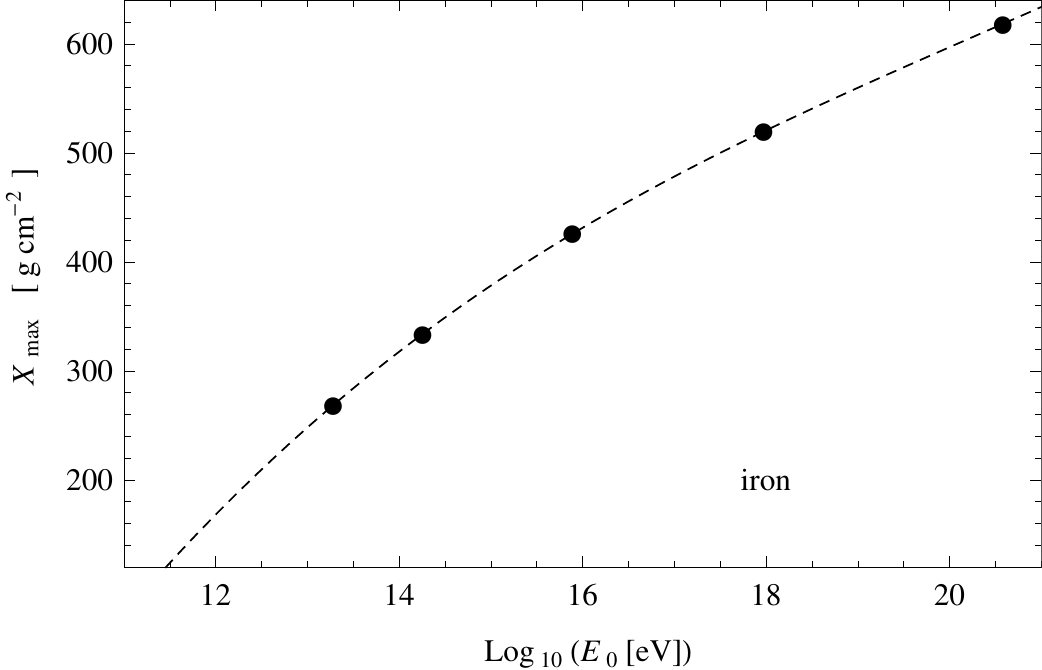}
\caption{The depth of maximum shower size as a function of energy of an iron primary for the situation with complete inelasticity (dots) and the corresponding cubic fit (dashed).}
\label{iroxmax}
\end{center}
\end{figure}

The hadronic cascade for an iron primary differs from the one for a proton primary by a smaller depth of first interaction and by a larger multiplicity in the first interaction. The iron-air cross section is about 2000 mb, see Fig. 54 of \cite{iron}. The corresponding interaction length is $\lambda_{\textrm{Fe-air}} \approx 12$ g cm$^{-2}$. The relatively small energy dependence of the iron-air cross section will only lead to a negligible difference of a few g cm$^{-2}$. According to the superposition model \cite{Gaisser} the multiplicity of a composite nucleus with atomic mass $A$ is equal to $A$ times the multiplicity of a proton with a $A$ times smaller energy:
\begin{equation}\label{55}
M_{ch} = 0.1 A \cdot \left( \frac{E_0}{A} \right)^{0.18} \ .
\end{equation}
For iron-air, $A=56$, this is $M_{ch} = 2.7 \cdot E_0^{0.18}$.  Since not all nucleons will participate in the same rate as a single proton, the latter should be multiplied by a factor smaller than unity. If this factor is (almost) independent on energy, the superposition model predicts a constant rate between the iron-air and proton-air multiplicity. This is indeed what is seen from QCD based models and from a color glass condensate approach, see Fig. 5 of \cite{Roesler} and Fig. 7 of \cite{Drescher}. On the basis of Fig. 5 of \cite{Roesler} we take in our model the iron-air multiplicity as
\begin{equation}\label{57}
M_{ch}=0.3  E_0^{0.18} \textrm{  ,     } \textrm{   } \textrm{  } M=0.45  E_0^{0.18} \ .
\end{equation}
In the absence of elasticity the iron-air multiplicity is only present in the first interaction. Except for these two adaptions the analysis is identical to the one for a primary proton. Starting with a primary iron with energy $E_0$ the energy of the particles after the first interaction is  
\begin{equation}\label{58}
E_1= \frac{E_0}{0.45 \cdot E_0^{0.18}} \approx 2.2 \cdot E_0^{0.82} \ .
\end{equation}
Without elasticity the subsequent interactions are governed by the pion multiplicity: 
\begin{equation}\label{59}
E_2= \frac{E_1}{0.15 \cdot E_1^{0.18}} \approx 6.7 \cdot E_1^{0.82} \ .
\end{equation}
Repeating the iteration we find for the energy per particle after $n$ interactions
\begin{equation}\label{60}
E_n=  6.7^{\alpha_{n-1}} \cdot E_1^{\beta_{n-1}} \ ,
\end{equation}
where $\alpha_n$ and $\beta_n$ are as defined in section 3.
Substitution of (\ref{58}) gives
\begin{equation}\label{61}
E_n=  6.7^{\alpha_{n-1}} \cdot 2.2^{\beta_{n-1}} \cdot E_0^{\beta_n} \ ,
\end{equation}
For the interaction lengths we have for the primary iron, $n=0$,
\begin{equation}\label{62}
\lambda_0 = \lambda_\textrm{Fe-air}(E_0) = 12 \ .
\end{equation}
The interaction lengths for the produced pions, $n \geq 1$, are as given before.
\begin{equation}\label{63}
\lambda_n = \lambda_{\pi \textrm{-air}}(E_n)= 200- 3.3 \ln (E_n) \ .
\end{equation}
With the substitution of (\ref{61}) this is 
\begin{equation}\label{64}
\lambda_n = 200-6.3 \alpha_{n-1} - 2.6\beta_{n-1} -3.3 \beta_n \ln (E_0) \ , \ \ \ n \geq 1 \ .
\end{equation}
The atmospheric depth of the $n_c$-th interaction, $n_c \geq 1$, becomes
\begin{eqnarray}\label{65}
X(n_c) =  200 n_c -188 -6.3 \sum_{n=1}^{n_c-1} \alpha_{n-1} \nonumber\\
- 2.6 \sum_{n=1}^{n_c-1} \beta_{n-1} - 3.3 \ln (E_0) \sum_{n=1}^{n_c-1}\beta_{n} \ .
\end{eqnarray}
For the decay length of the charged pions after the $n_c$-th interaction we now find
\begin{equation}\label{66}
c \tau \gamma [\textrm{km}] \approx 5.6 \cdot 10^{-11} \cdot 6.7^{\alpha_{n_c-1}} \cdot 2.2^{\beta_{n_c-1}} \cdot E_0^{\beta_{n_c}} \ . 
\end{equation}
So, for the depth of the decay of the pions we obtain the following equation: 
\begin{equation}\label{67}
5.6 \cdot 10^{-11} \cdot 6.7^{\alpha_{n_c-1}} \cdot 2.2^{\beta_{n_c-1}} \cdot E_0^{\beta_{n_c}} =  4 \ln \left( \frac{X(n_c+1)}{X(n_c)} \right) \ . 
\end{equation}
As before, we substitute integer values for $n_c$ and solve numerically for the primary energy $E_0$. Results of interest are shown in table 4, 5 and 6, the equivalents of table 1, 2 and 3.
\\  
\begin{table}[htbp]
\renewcommand{\arraystretch}{1.5}
\setlength{\tabcolsep}{5pt}
\begin{tabular}{|c |c |c |c |c |c |c |c |c |}\hline 
$n_c$ & $E_0$ [eV] & $X(n_c)$ & $X(n_c+1)$ & $h(n_c)$ [km] & $\Delta h$ [km] & $\xi_c^{\pi}$ [eV]   \\ \hline 
$1$ & $1.9 \cdot 10^{13}$ & $12$ & $127$ & $35.6$ & $18.9 $ & $1.7 \cdot 10^{11}$  \\ 
$2$ & $1.8 \cdot 10^{14}$ & $121$ & $239$ & $17.2$ & $5.5 $ & $4.9 \cdot 10^{10}$  \\ 
$3$ & $7.7 \cdot 10^{15}$ & $221$ & $341$ & $12.3$ & $3.5 $ & $3.1 \cdot 10^{10}$  \\ 
$4$ & $9.4 \cdot 10^{17}$ & $309$ & $430$ & $9.6$ & $2.6 $ & $2.4 \cdot 10^{10}$  \\
$5$ & $3.8 \cdot 10^{20}$ & $380$ & $502$ & $8.0$ & $2.2 $ & $2.0 \cdot 10^{10}$  \\ \hline
\end{tabular}
\caption{Characteristics of hadronic cascades for a primary iron.}
\end{table}
\begin{table}[htbp]
\renewcommand{\arraystretch}{1.5}
\setlength{\tabcolsep}{5pt}
\begin{tabular}{|c |c |c |c |c |c |c |c |c |}\hline 
$n_c$ & $E_0$ [eV] & $\lambda_0$ & $\lambda_1$ & $\lambda_2$ & $\lambda_3$ & $\lambda_4$ & $\bar \lambda$ \\ \hline
$1$ & $1.9 \cdot 10^{13}$ & $12$ & -- & -- & -- & -- & 12 \\ 
$2$ & $1.8 \cdot 10^{14}$ & $12$ & $109$ & -- & -- & -- & 60 \\ 
$3$ & $7.7 \cdot 10^{15}$ & $12$ & $98$ & $110$ & -- & -- & 74 \\ 
$4$ & $9.4 \cdot 10^{17}$ & $12$ & $85$ & $100$ & $111$ & -- & 77 \\ 
$5$ & $3.8 \cdot 10^{20}$ & $12$ & $69$ & $86$ & $101$ & $112$ & 76 \\ \hline
\end{tabular}
\caption{Subsequent interaction lengths in hadronic cascades for a primary iron. The $\lambda_n$ and the cascade average $\bar \lambda$ are in g cm$^{-2}$.}
\end{table}
\begin{table}[htbp]
\renewcommand{\arraystretch}{1.5}
\setlength{\tabcolsep}{5pt}
\begin{tabular}{|c |c |c |c |c |c |c |c |c |}\hline 
$n_c$ & $E_0$ [eV] & $M_0$ & $M_1$ & $M_2$ & $M_3$ & $M_4$ & $\bar M$ \\ \hline
$1$ & $1.9 \cdot 10^{13}$ & $110$ & -- & -- & -- & -- & $110$ \\ 
$2$ & $1.8 \cdot 10^{14}$ & $165$ & $22$ & -- & -- & -- & $60$\\ 
$3$ & $7.7 \cdot 10^{15}$ & $325$ & $38$ & $20$ & -- & -- & $63$ \\ 
$4$ & $9.4 \cdot 10^{17}$ & $773$ & $78$ & $36$ & $19$ & -- & $80$ \\ 
$5$ & $3.8 \cdot 10^{20}$ & $2274$ & $189$ & $73$ & $34$ & $18$ & $114$ \\ \hline
\end{tabular}
\caption{Subsequent multiplicities and their geometric mean in hadronic cascades for a primary proton.}
\end{table}
\\
For the energy entries in table 4 the maximum number of electrons and muons for an iron initiated air shower is calculated in a similar way as for the proton initiated air showers. The resulting curves, see Fig. \ref{irosizes}, are practically identical to the ones for a proton primary in Fig. \ref{sizes}. A linear fit, see the dashed lines in Fig. \ref{irosizes}, for the maximum number of electrons ($\pm$) and number of muons yields
\begin{equation}\label{67a}
N_\textrm{e}  \approx 0.59 \cdot ( E_0 [\textrm{GeV}] )^{1.010} 
\end{equation}
and
\begin{equation}\label{68}
N_\mu \approx 0.0055 \cdot ( E_0 [\textrm{GeV}] )^{1.016} 
\end{equation}
respectively. As for proton primaries, the values of $N_e + 25 N_\mu$ for different values of primary iron energy $E_0$ are close to the ones obtained with Monte Carlo models, see Fig. 3 of \cite{Glasmacher}.
\\ \\
Also the depth of maximum shower size for an iron primary is calculated in a similar way as for a proton primary. The result is shown in Fig. \ref{iroxmax}. 
\\ \\
Both the smaller interaction length and the larger multiplicity have reduced the depth of maximum shower size with respect to a proton initiated shower. A cubic fit, see the dashed line in Fig. \ref{iroxmax}, yields
\begin{equation}\label{70}
X_{\textrm{max,Fe}} \approx  -2116+329.4\lg E_0-14.43(\lg E_0 )^2+0.237 (\lg E_0)^3 \ ,
\end{equation}
where the subscript Fe identifies the primary particle. As for proton showers the curve is almost linear for energies above $10^{14}$ eV:
\begin{equation}\label{71}
X_{\textrm{max,Fe}} \approx  -305+45\lg E_0 \ .
\end{equation}
As for the proton an analytical estimate can be obtained by considering solely the first generation of $\gamma$'s:
\begin{equation}\label{72}
X_{\textrm{max,Fe}} \approx \lambda_{\textrm{Fe-air}} + X_{\textrm{max},\gamma}(E_0/(2M))\approx -659+69.8 \lg E_0 \ .
\end{equation}
Also here the actual elongation rate is substantially smaller when further generations are taken into account.

\section{Energy distribution}
In the Heitler-Matthews model as well as in the present extension the energy of the interaction is assumed to  be equally divided over the secondary particles. In reality the distribution of energy is highly inhomogeneous. This is observed in proton-proton collisions, see Fig. 48 of \cite{Alt}. For proton-air collisions it is predicted by Monte Carlo models, see Fig. 6 of \cite{Rebel}. Many secondaries obtain a small part of the energy while a few particles obtain a larger part. The elasticity effect, where a substantial part of the energy is taken by the leading particle, can be regarded as the most profound manifestation of the inhomogeneous energy distribution. As will be argued both the elasticity and the inhomogeneous energy distribution over the non-leading secondaries increase the depth of maximum shower size. In our opinion the equal division of energy is therefore responsible for a substantial part of the discrepancy with respect to shower simulators which have the inhomogeneous energy distribution incorporated. To give some foundation to the idea we consider the following relation \cite{Pajares}:
\begin{equation}\label{80}
\frac{\Delta X_{\textrm{max}}}{X_{\textrm{max}}} \approx -\frac{1}{2}\frac{\Delta M}{M}-\frac{1}{10}\frac{\Delta \kappa}{\kappa} \ .
\end{equation} 
In this relation $\Delta X_{\textrm{max}}$ is the shift of the depth of maximum shower size, $M$ is the multiplicity and $\kappa$ is the inelasticity. For protons and pions the inelasticity is roughly about $\frac{2}{3}$, see figure 6 of \cite{Alvarez}. Starting from the inelastic situation,$\kappa =1$, the change in inelasticity is $\Delta \kappa = -\frac{1}{3}$. According to the second part in the right hand side of Eq. (\ref{80}) this corresponds to a shift
\begin{equation}\label{81}
\Delta X_{\textrm{max}} \approx \frac{1}{30} \cdot X_{\textrm{max}}\ .
\end{equation}
That is, about one seventh of the discrepancy can be explained by elasticity. 
\\ \\
Also the inhomogeneous energy distribution over the non-leading secondaries increase the depth of maximum shower size. Among the secondary charged pions there will be a few with relatively large energy who penetrate deeper into the atmosphere thereby contributing to the depth of maximum shower size in a similar way as elasticity does. At the same time there will be many secondary charged pions with energies so low that they will decay before they reach final generation of the cascade. In effect this reduces the cascade mean multiplicity. From the first part of the right hand side of Eq. (\ref{80}) it follows that a substantial contribution to the shift $\Delta X$ can be expected. Since the effective reduction of multiplicity will be larger for large multiplicities and thus for large energies, the relative shift will be larger for large energies. As a consequence the inhomogeneous energy distribution over the non-leading secondaries does increase the depth of maximum shower size as well as the elongation rate. To quantify these statements we consider the first generation prediction (\ref{46}) and investigate its sensitivity for the multiplicity. A change in the multiplicity $M=q \cdot E^p$ can be obtained by a change of $q$ and a change of $p$. From the derivative of (\ref{46}) with respect to $q$ we obtain
\begin{equation}\label{83}
\frac{\partial X_{\textrm{max,p}}}{\partial q} \approx -\frac{37}{q}\ . 
\end{equation}
A decrease of the multiplicity by a decrease of $q$ leads to an increase of $X_{\textrm{max,p}}$ independent of the energy $E_0$. Alternatively, a decrease of $q$ increases solely the absolute level of $X_{\textrm{max,p}}$. Variations of $q$ do not affect the elongation rate. From the derivative of (\ref{46}) with respect to $p$ we obtain
\begin{equation}\label{84}
\frac{\partial X_{\textrm{max,p}}}{\partial p} \approx -85 \lg E_0  \ . 
\end{equation}
Here a decrease of the multiplicity by a decrease of $p$ leads to an increase of $X_{\textrm{max,p}}$ proportional to  $\lg E_0$. This is slightly suppressed by second and further generations. The important conclusion however is that the variation of $p$ affects both the absolute value of $X_{\textrm{max,p}}$ and the elongation rate. 
\\ \\
Since it makes a difference whether the multiplicity is varied by $q$ or by the power $p$ it is better to distinguish the sensitivity of $X_{\textrm{max,p}}$ for it. That is, instead of expressing $\Delta X_{\textrm{max}} /X_{\textrm{max}}$ in terms of $\Delta M / M$ we express it in terms of $\Delta p / p$ and $\Delta q / q$. The substitution of (\ref{83}) and (\ref{84}) in
\begin{equation}\label{85}
\frac{\Delta X_{\textrm{max,p}}}{ X_{\textrm{max,p}}} = \frac {1}{X_{\textrm{max,p}}} \frac{\partial X_{\textrm{max,p}}}{\partial p} \Delta p +\frac {1}{X_{\textrm{max,p}}} \frac{\partial X_{\textrm{max,p}}}{\partial q} \Delta q  \ , 
\end{equation}
leads to
\begin{equation}\label{86}
\frac{\Delta X_{\textrm{max,p}}}{ X_{\textrm{max,p}}} = -\frac {85 \lg E_0}{X_{\textrm{max,p}}} \Delta p - \frac {37}{X_{\textrm{max,p}}}  \frac{\Delta q}{q}  \ . 
\end{equation}
From $M=qE^p$ it follows that $ \Delta M / M = \Delta q /q$ and $ \Delta M / M = 2.3 \lg E_0 \Delta p$. So, if the multiplicity is varied solely by either $q$ or $p$, we can write (\ref{86}) as
\begin{equation}\label{87}
\frac{\Delta X_{\textrm{max,p}}}{ X_{\textrm{max,p}}} =  - \frac {37}{X_{\textrm{max,p}}}  \frac{\Delta M}{M}  \ . 
\end{equation}
For a depth of maximum of, say, $X_{\textrm{max,p}}\approx 550$ g cm$^{-2}$ the latter is of the order
\begin{equation}\label{88}
\frac{\Delta X_{\textrm{max,p}}}{ X_{\textrm{max,p}}} \approx  - \frac {1}{15}  \frac{\Delta M}{M}  \ . 
\end{equation}
The latter suggests that the factor $\frac{1}{2}$ in (\ref{80}) is an overestimation of the sensitivity of the depth of maximum for variations of the multiplicity.

\section{Conclusions and discussion}
\begin{figure}[htbp]
\begin{center}
\includegraphics[width=11.0cm]{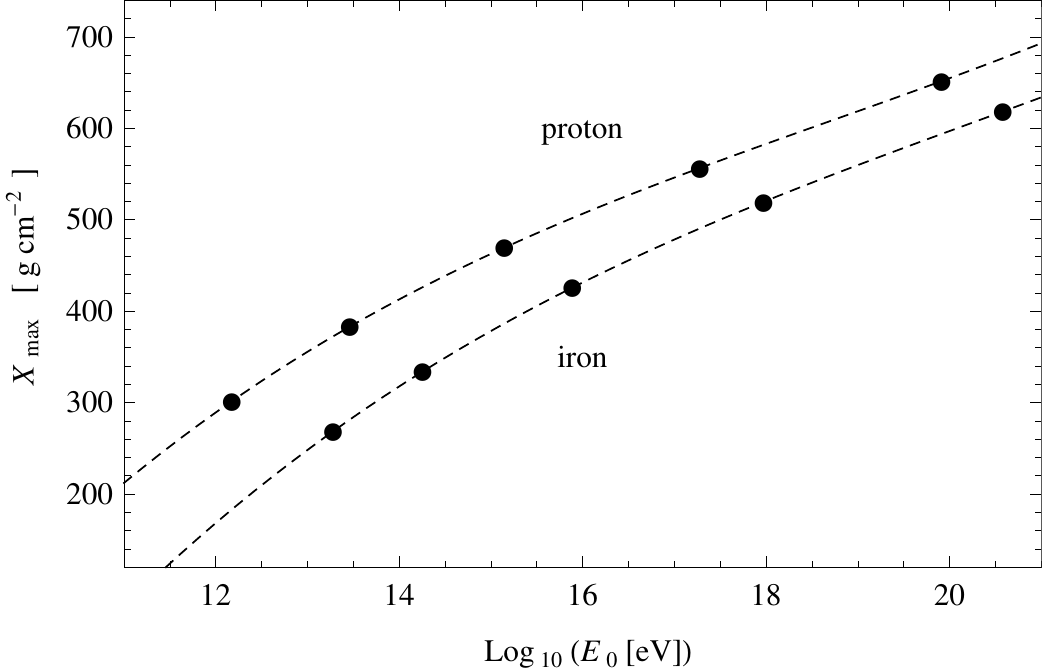}
\caption{The depth of maximum shower size as a function of energy for proton and iron primaries for the situation with complete inelasticity (dots) and the corresponding cubic fits (dashed).}
\label{protonandiron}
\end{center}
\end{figure}

Hadronic cascades in cosmic air showers are analyzed by means of a Heitler-Matthews model extended to all generations of pions. For all the predictions a multiplicity and interaction length is applied parameterized for energy. It is argued that the thickness of the interaction layers should be taken equal to the interaction length and not a fraction $\ln 2$ of it. Although this increases the prediction for the depth of maximum shower size with a few tens of g cm$^{-2}$, the value for $X_{\textrm{max}}$ still is too small in comparison with Monte Carlo simulations. It is shown that an analysis based on all the generations in the hadronic cascade does lead to smaller elongation rates than an analysis solely based on the first generation of $\gamma$ showers. The agreement of the latter with the Monte Carlo prediction for the elongation rate can therefore be considered as coincidental. The depth of maximum curves for proton and iron primaries as predicted by the present model are both shown in Fig. \ref{protonandiron}. 
\\ \\
The curves do show some similarities to the ones resulting from Monte Carlo simulations \cite{Knapp, Rebel,Horandel2,Auger}: they are almost linear for energies larger than $10^{14}$ eV, at low energies the elongation rate tends downwards for increasing energies and the separation between the proton and iron curve is in agreement with Monte Carlo simulations. Also the eleongation rate for iron being larger than for proton, and the corresponding decrease of the separation for increasing energy is in agreement with Monte Carlo simulations. From the present model this can be understood as follows. The smaller interaction length causes a lower level for the iron curve. If the iron and proton multiplicities would be identical the curves would run parallel. The larger iron multiplicity has two effects. It reduces the energy of the first generation of subshowers and a larger initial energy is required for the shower to survive the same number of interactions. The first effect decreases the level of the iron curve, while the second effect shifts the curve to larger energies. Since the second and further generations are governed by pion multiplicities, the cascade average multiplicity for iron tends towards the one for a proton primary. As a consequence the separation between corresponding dots in Fig. \ref{protonandiron} becomes smaller for larger energies, resulting in an iron curve that tends to the proton curve for increasing energy.
\\ \\ 
Despite the agreements with Monte Carlo simulations there are two important differences: the absolute levels of the curves are too small and the elongation rates are too small. For a proton and iron primary the depth of maximum shower size as predicted by Monte Carlo models, see for instance the right panel of Fig. 10 of \cite{Knapp}, Fig. 13 of \cite{Rebel} or Fig. 9 of \cite{Horandel2}, is about 
\begin{equation}\label{91}
X_{\textrm{max,p}} \approx -310 + 58 \lg E_0 
\end{equation}
and 
\begin{equation}\label{92}
X_{\textrm{max,Fe}} \approx -580 + 67 \lg E_0  . \ 
\end{equation}
respectively. The comparison with the present model predictions (\ref{42}) and (\ref{71}) learns that the Monte Carlo predictions are larger by about 100 g/cm$^2$ at $10^{14}$ eV up to more than 150 g/cm$^2$ at $10^{20}$ eV. In our opinion a substantial part of the discrepancy may be caused by the homogeneous energy distribution of the secondaries in the discrete model. An inhomogeneous energy distribution will effectively reduce the multiplicity. As argued in the previous section a decrease of the power $p$ in the relation $M=qE^p$ does increase both the depth of maximum and the elongation rate. Both are needed to explain the discrepancy with Monte Carlo simulations. The fact that the present model prediction for the elongation rate also needs to be increased in order to match with Monte Carlo simulations can even be regarded as a support of the present inclusion of all the generations of electromagnetic showers. 
\\ \\
Of course one can think of several other possible contributions to the discrepancy. For instance, the present model is discrete. The interactions occur at discrete intervals and the stopping occurs rigidly when the decay length is half the interaction length. It cannot be excluded that the inclusion of statistics in these processes will affect the prediction for depth of maximum. This is beyond the scope of the present discrete model. 
\\ \\

{\Large \textbf{Acknowledgements}}\\
I am grateful to the reviewers for their useful comments. I wish to thank Dr. J.J.M. Steijger for his detailed and useful comments on an earlier draft of this paper. I wish to thank prof. J.W. van Holten and Prof. B. van Eijk for their useful comments and encouragement. I wish to thank Nikhef for its hospitality. The work is supported by a grant from NWO (Netherlands Organization for Scientific Research).

\newpage

\bibliography{References}

\end{document}